\documentclass[prd,twocolumn,superscriptaddress,altaffilletter,amssymb,showpacs,epsfig,nofootinbib]{revtex4}
\usepackage[dvips]{graphicx}
\usepackage{amsmath}
\newcommand{\be}{\begin{equation}}
\newcommand{\ee}{\end{equation}}
\newcommand{\ben}{\begin{eqnarray}}
\newcommand{\een}{\end{eqnarray}}
\newcommand{\n}{\label}

\newcommand{\no}{\noindent}

\newcommand{\ga}{\gamma}

\begin{document}

\title{Crossing the phantom barrier with purely kinetic multiple k-essence}
\author{Luis P. Chimento}\email{chimento@df.uba.ar}\affiliation{Dpto. de F\'\i sica, Facultad de Ciencias Exactas y Naturales,
Universidad de Buenos Aires, Ciudad Universitaria
Pabell\'on I, 1428 Buenos Aires, Argentina}
\author{Ruth Lazkoz}
\email{ruth.lazkoz@ehu.es}
\affiliation{Fisika Teorikoa, Zientzia eta Teknologia Fakultatea, Euskal Herriko Unibertsitatea, 644 Posta Kutxatila, 48080 Bilbao, Spain}
\begin{abstract}
We consider multiple k-essence sources and obtain the conditions their kinetic functions must satisfy so that purely kinetic k-essences lead to models with phantom barrier crossing. After that, we show that polynomial kinetic functions allow the integration of the dynamical equations determining the geometry and the k-fields. The models thus obtained have accelerated expansion  at late cosmic time, representing  universes ending at a finite time with a big rip singularity. In addition, these models begin to evolve from an initial singularity, so they have a finite time span.
 \end{abstract}
\maketitle
\section{Introduction}Over the last years we have witnesses the steady rise of a new standard cosmology according to which our universe is currently spatially flat, has a low matter content, and has recently began to expand acceleratedly.
The strongest support for accelerated expansion is provided by observations of distant supernovae \cite{obs1,obs2}, but it is also independently supported by cosmic microwave background \cite{obs3,obs4,obs5}
and large scale structure observational information \cite{obs6}.

The simplest explanation for dark energy is a cosmological constant, but the fine-tuning of its value needed according to theoretical foundations has motivated a painstaking pursue to find more satisfactory explanations. There is a broad consensus that dark energy must be dynamical, and the most popular models resort to scalar fields.

Up to quite short ago, many observational results had been collected which pointed in the direction  that the dark energy equation of state parameter $w\equiv\gamma-1=p/\rho$ has recently crossed the $w=-1$ barrier (see \cite{cabre} for very recent conclusions along theses lines using CMB and large structure data).  Nevertheless, this apparently settled issue has been questioned  by some recent studies using newest SN data. Those works suggest models with phantom divide crossing might misleadingly provide better fits just because of underlying systematic errors in observations \cite{vanish}. In contrast, other researchers  conclude that all classes of dark energy models are comfortably allowed \cite{barger} by those very observations.  At this stage there is not but a few of studies with seem to rule out the necessity of the existence of a $w<0$ epoch, and to make things even more debatable one cannot forget results have strong dependence on the chosen parametrization of $w$ in terms of redshift \cite{doran}, so there is neither full agreement nor disagreement  about the observational preference of the phantom divide crossing. This being the state of the art, it seems further advances in this area will be necessary  one can ascertain which is the correct opinion, so continuing to pursue  models which do the crossing seems still justified.

Given that the crossing cannot be implemented with a single field (no matter if its kinetic term is canonical or not \cite{theoretical}) there has been quite a lot of activity oriented toward engineering
possible ways to realize such crossing \cite{caldor,brane,string,theoretical,hu,so,high,vector,hybrid,scalar,per,hess,quintom,chap,
dscross1,dscross2,binf,rg}.
A possibility which has been hardly explored is that of scalar fields with non-canonical
kinetic terms, leading to a type of dark generically dubbed k-essence \cite{pioneer, k-essence}. Here we make explicit detailed constructions  stemming from very general assumptions which show once again the versatility of k-essence dark energy. Our models are constructed using a particularly simple class
of k-essence models in which the Lagrangian does not depend on the fields themselves but only on their derivatives. These class of k-essence models were the pioneering ones \cite{pioneer}, and
have afterward received some attention as possible candidates for the unification of dark energy and dark matter \cite{feinstein}-\cite{chif}. Precisely a class of Chaplygin cosmologies (the so called modified ones) \cite{chimento} are in fact
kinetic k-essence models.

This letter
is structured as follows. After presenting the dark energy Lagrangian defining the multiple scalar field sources of our geometries, we outline the evolution equations of those scalar fields and the geometric degrees of freedom. We then concentrate on two-field configurations and search for general necessary conditions for the crossing to occur. After that, we select kinetic functions satisfying such conditions, which at the same will
allow obtaining closed formed expressions for the geometry and the k-field.  We close with a section on conclusions and future prospects.

\section{Multiple k-essence}
\subsection{Preliminaries}
Our investigation is  devoted to the cosmological evolution of multiple scalar k-fields $\phi_i$ with $i=1,\dots, N$ when they are minimally coupled to gravity and driven by the individual potentials $V_i(\phi_i)$. For this description we resort to the Lagrangian
\ben
S_{tot}=S_{g}+S_{de}\\
\een
where
\ben
&&S_{g}=\int d^4x\sqrt{-g}\,\frac{R}{6},\qquad\\
&&S_{de}=-\int d^4x\sqrt{-g}\,\sum_{i=1}^N V_i(\phi_i)F_i(X_i).\qquad
\een 
The former is the gravitational action and the later corresponds to the k-essence action with non-canonical kinetic terms and containing derivatives of no higher order than first. Besides, we denote
\ben
X_i\equiv g^{\mu\nu}\phi_{i\mu}\phi_{i\nu},
\een
where $\phi_{i\mu}=\partial_{\mu}\phi_{i}$ .
Variation of the action $S_{de}$ with
respect to the metric gives the energy momentum tensor of the dark energy:
\ben
T_{\mu\nu}=\sum_{i=1}^NV_i[{F_i'}\phi_{i\mu}\phi_{i\nu}-g_{\mu\nu}F_i]\label{tmunu}\,.
\een
Here and throughout primes denote differentiation with respect to the argument. Interpreting the sources of geometry as perfect fluids their energy density and pressure will be given by
\ben
&&\rho_i=V_i[F_i-2X_i F_i'], \qquad  p_i=-V_iF_i.
\een

Considering a spatially flat, homogeneous and isotropic spacetime with   scale factor $a(t)$ and  Hubble factor $H\equiv \dot a/a$, and  differentiating $S_{g}$ with respect the metric, we will obtain the Einstein equations, which read
\ben
3H^2&=&\sum_{i=1}^NV_i[F_i-2X_i F_i'],\quad\label{frie}\\
\dot H&=&\sum_{i=1}^NV_iX_iF_i'\label{ray}
\een
where
\be
X_i=-\dot\phi_i^2\label{grad}.
\ee


We want to concentrate on purely kinetic models so we will simply take constant potentials $V_i>0$. In addition, we will consider our cosmologies are driven by $N$ non-interacting k-essences, so there will be $N$ independent conservation equations. The assumption that the potentials are constant allows to write  the first integral of the k-field equations \cite{chimento}, so those conservation equations become
\ben
a^3\dot\phi_iF_i'=\frac{b_i}{V_i}\label{fi1}
\een
with $b_i$ 
some real constants. 

In the next subsection we present a very general choice for the kinetic functions $F_i$ 
which generates a large set of cosmologies which cross the phantom barrier. This represents a considerable progress in this area compared with previous works in the literature where only a few particular examples were obtained after a standard reconstruction process following ad hoc choices of the geometry.

\subsection{General kinetic functions}

Let us try an extract as much information as we can on the general conditions the kinetic functions must obey so that our models can do the passage from a conventional to a phantom era. The transition will be possible provided the sign of $\dot H$ is allowed to change (the conventional epoch is characterized by $\dot H<0$, whereas the phantom one occurs when
$\dot H>0$). Under the specifications made above one, and using Eqs. (\ref{ray}-\ref{fi1}) one can always write
\be
\dot H=-\frac{1}{a^3}\sum_{i=1}^N b_i\dot\phi_i.
\ee
Thus, we see a  necessary condition for the crossing is that not all the products 
$b_i\dot\phi_i$ have the same sign.
If we now use Eq. (\ref{ray}), we see such condition will be satisfied provided
not all the signs of the quantities  $F_i'$ coincide. Having derived this necessary condition, let us now move on to polynomial models.

\subsection{Polynomial kinetic functions}

The evolution of  k-essence models with polynomial kinetic functions has been investigated in \cite{feinstein,chimento,aguirre,chif}. Those are the only known kinetic functions which lead to 
a scale factor with a power-law dependence on the scalar field without giving  constant gradients of the $\phi_i$ k-fields.
For the various reasons we have discussed so far we take as our starting point
\be
F_i=\pm(-X_i)^{{\gamma_i}/{(2(\gamma_i-1))}}\label{kinfun},
\ee
which leads to
\be
\rho_i=\frac{V_iF_i}{1-\gamma_i},
\ee
with $\ga_i$ some constants. The latter suggests the cases in which any of the $\gamma_i$ is equal to $1$ are ill-defined.

Substituting Eqs. (\ref{grad},\ref{kinfun}) into Eq. (\ref{fi1}) and integrating one gets
$F_i\propto {a^{-3\gamma_i}}$; and consequently one can write
\be
\n{ri}
\rho_i=\frac{c_i}{a^{3\gamma_i}},
\ee

\no where $c_i$ are constant.
%

To make a closer contact with related works in the literature  we will take $N=2$ in what follows. This restriction, however, is not too significant because one still has the freedom generated by the biparametric character of the models we are going to investigate.
The above choices lead to
\ben
3H^2=
\frac{c_1}{a^{3\gamma_1}}+\frac{c_2}{a^{3\gamma_2}},
\label{frie}\\
-2\dot H=\frac{\gamma_1c_1}{a^{3\gamma_1}}+\frac{\gamma_2c_2}{a^{3\gamma_2}}.
\n{.H}
\een


Remember that the conventional epoch is characterized by negative $\dot H<0$, whereas the phantom one occurs when that quantity is positive. Therefore in an expanding universe the transition between {\it non-phantom and phantom regimes} will be possible at $t_c$ where $a_c=a(t_c)$ provided that $\dot H(a_c)=0$ and $(d\dot H/dt)\vert_{t=t_c}>0$. Restricting ourselves to expanding universes then the second condition for the crossing actually becomes $(d\dot H/da)\vert_{a=a_c}>0$. From Eq. (\ref{.H}) we conclude
\ben
\n{con}
a_c^{3(\gamma_1-\gamma_2)}=-\frac{c_1\gamma_1}{c_2\gamma_2},\\
\frac{d\dot H}{da}=\frac{3}{2a}\left[\frac{c_1\gamma_1^2}{a^{3\gamma_1}}+\frac{c_2\gamma_2^2}{a^{3\gamma_2}}\right],
\een
and these expressions give rise to two cases:
\begin{itemize}
\item Case i) It corresponds to $c_1>0$ and $c_2>0$ with barotropic indexes $\gamma_1$, $\gamma_2$ having different signs. 
In this situation the barrier crossing will always be possible because the derivative $(d\dot H/da)\vert_{a_c}$ is positive definite for any value of the scale factor. \\
\item Case ii) When the signs of  $c_1$ y $c_2$ are different and the barotropic indexes $\ga_1$, $\ga_2$ have the same sign then  $a_c$ exists as well. In this case we have
\be
\n{ii}
\left(\frac{d\dot H}{da}\right)\bigg\vert_{a=a_c}=\frac{3c_2\gamma_2(\gamma_2-\gamma_1)}{2a_c^{3\gamma_2+1}}
\ee
so that if we take $c_1<0$ and $c_2>0$, the derivative above will be positive under two circumstances: 
\begin{description}
 \item Subcase a) $\gamma_1>0$ and $\gamma_2>0$ with $\gamma_2>\gamma_1$,\\

 \item Subcase b) $\gamma_1<0$ and $\gamma_2<0$ with $\gamma_2<\gamma_1$.\end{description}
In both subcases from the Friedmann equation one deduces that the scale factor has an extremum at $a=a_e$, so from Eq. (\ref{con}) we get
\be
\n{et}
\left(\frac{a_c}{a_e}\right)^{3(\gamma_1-\gamma_2)}=\frac{\gamma_1}{\gamma_2},
\ee
but on the other hand, on the extremum
\be
\n{e}
\dot H_e=\frac{c_2(\gamma_1-\gamma_2)}{2a^{3\gamma_2}_e}=\frac{\ddot a_e}{a_e}.
\ee
In subcase a) we see $a_e$ is a maximum of the scale factor, i.e. $a_e=a_{\rm max}$, so $a_c<a_{max}$, whereas from Eq. (\ref{et}) we conclude $\gamma_1>\gamma_2$, which contradicts the hypothesis of subcase a). Likewise, in subcase b) we see $a_e$ is a minimum of the scale factor, i.e. $a_e=a_{\rm min}$, so $a_c>a_{min}$, whereas from Eq. (\ref{et}) $\gamma_1<\gamma_2$ follows, and it contradicts the  hypothesis of subcase b).

\end{itemize}

Summarizing, case i) is the only suitable one for our purposes. In order to find the exact general solution of the Friedmann equation (\ref{frie}), let us introduce now a new time variable $\tau$ and a new time-dependent variable $s(\tau)$,  
\ben
 &&s=\omega a^{{3(\gamma_1 -\gamma_2)}/{2}}\\
&&\tau=\frac{\sqrt{3}\,(\gamma_1-\ga_2)}{2} \int a^{{-3\gamma_2}/{2}}dt\label{tautot}.\,
\een 
where $\omega^2=c_2$ and without loss of generality we have set $c_1=1$. After some simple steps one can recast the Friedmann equation in the form
\ben
{s'^2}-\omega^2({1+s^2})=0.
\een
The latter can be readily integrated to give 
$
s=\sinh \omega \Delta\tau
$, so that
\be
\n{a}
a=\left[\frac{1}{\omega}\sinh\left(\omega \Delta\tau\right) \right]^{2/(3(\gamma_1-\gamma_2))},
\ee
where $\Delta\tau=\tau-\tau_0$ with $\tau_0$ an arbitrary integration constant.
Inserting the latter into (\ref{tautot}) one gets 
\ben
\n{t}
&&\hspace{-0.5cm}t=\frac{2}{\sqrt{3}\,(\gamma_1-\gamma_2)}\int \left[\frac{1}{\omega}\sinh\left(\omega \Delta\tau\right) \right] ^{\gamma_2/(\gamma_1-\gamma_2)}d\tau,\
\een
and this solves completely the evolution of this kind of universes. 

It is easy to combine our expressions for $a(\tau)$ and $t(\tau)$ to obtain the final and initial behavior of the scale factor. 
Assuming $\ga_1>0$ and $\ga_2<0$, in the large $\tau$ limit one can approximate Eq. (\ref{t}) by
$t_f-t\approx e^{w\Delta\tau\gamma_2/(\gamma_2-\gamma_1)}$, so we conclude that $\tau=\infty$ corresponds to the final finite time $t_f$ and the scale factor (\ref{a}) can be approximated by $a\approx (t_0-t)^{2/3\ga_2}$. This means the blow up of the scale factor and its derivatives is reached suddenly at the finite cosmic time $t_f$, so we can ascertain our models have a final big rip.
On the other hand, in the short $\Delta\tau$ limit one can approximate Eq. (\ref{t}) by $t-t_i\approx\Delta\tau^{\ga_1/(\ga_2-\gamma_1)}$, so we conclude that $\Delta\tau=0$ corresponds to the initial finite time $t_i$ and the scale factor (\ref{a}) can be approximated by $a\approx (t-t_i)^{2/3\ga_1}$. This means models of this  kind  begin to evolve from an initial singularity at the finite time $t_i$ and end at $t_f$ with a finite time span. 

Interestingly, our models admit an interpretation in terms of the of two barotropic fluids interacting through the geometry only. One could then think of an effective barotropic index
$\gamma_{\rm eff}\equiv w_{\rm eff}+1=-2\dot H/3H^2$, which in our case reads
\be
\gamma_{\rm eff}=\frac{\gamma_1+\gamma_2\sinh^2 \omega \Delta\tau}{\cosh^2 \omega\Delta\tau}\,,
\ee
so as $t$ grows the model interpolates between $\gamma_1$ and $\gamma_2$, which is an alternative way to show the passage between the conventional and phantom regimes occurs indeed. In particular it is possible to obtain examples of universes which transit between a matter dominated era and a final phantom dark energy dominated scenario.

Finally, from Eqs. (\ref{kinfun})-(\ref{ri}), (\ref{a}) and (\ref{t}) explicit expressions for the k-fields can be obtained too.

\section{Conclusions and future prospects}
This paper is motivated by a controversial aspect of dark energy models which has received some attention recently: Has the dark energy equation of state transited recently from a conventional to a phantom character? Leaving aside the question of whether that is the case or not, which seems to be not settled we turn our attention to a possible way to construct models of that sort with the hope they can serve both as models for that crossing in the case the observational debate ends up showing they are in fact the best models, and with
the additional intention that they can help us deepen our understanding of  the kinematics of cosmological models with more than one component.

The use of multiple purely kinetic k-essence as a source of cosmological models proves once more to very fruitful. That choice allows one to construct a large set of explicit exact models which do the passage between the typical accelerated regime ($\dot H<0$)  to a super-accelerated or phantom ($\dot H>0$) regime.

The initial output is just the way the kinetic functions of the k-essence fluids depend on the gradients of the k-fields. This represents a more novel procedure than the alternative of performing a reconstruction process after giving as input ad hoc expressions for $H(t)$ and $a(t)$.

Interestingly, our models are exact solutions with a final big rip which do not belong to the familiar class of power-law cosmologies, so we can say they stand out from the crowd of exact solutions in this area.

Once again, kinetic k-essence has proved to be useful for depicting cosmic acceleration, just recall it can either play the role of the inflation or alternatively be interpreted as a component for the unification of dark matter and dark energy. This time, without leaving the realm of cosmologies with accelerated expansion, kinetic k-essence  allows building a filling for the universe which makes it cross the phantom divide 
\section*{Acknowledgments}
We are grateful to J.M. Aguirregabiria for valuable suggestions.
R.L. is supported by the Spanish Ministry of Science and Education through the RyC program, and research grants FIS2004-01626 and FIS2005-01181. L.P.C. is partially funded by the University of Buenos Aires  under
project X224.

\end{document}